\documentclass[conference,10pt]{IEEEtran}
\IEEEoverridecommandlockouts

\usepackage{epsfig,rotating,setspace,latexsym,amsmath,epsf,amssymb,amsfonts,bm,theorem,cite,algorithm,graphicx,epsf,authblk,epstopdf,color,algpseudocode,bbm}

\newtheorem{remark}{Remark}

\newtheorem{theorem}{Theorem}
\newtheorem{lemma}{Lemma}
\newenvironment{Proof}[1]{\medskip\par\noindent{\bf Proof:\,}\,#1}{{\mbox{\,$\blacksquare$}\par}}

\allowdisplaybreaks

\begin{document}

\title{Timely Estimation Using Coded Quantized Samples\thanks{This work was supported in part by the U.S. National Science Foundation under Grants CCF-0939370 and CCF-1513915.}}

\author[1]{Ahmed Arafa}
\author[2]{Karim Banawan}
\author[3]{Karim G. Seddik}
\author[4]{H. Vincent Poor\vspace{-.1in}}
\affil[1]{\normalsize Department of Electrical and Computer Engineering, University of North Carolina at Charlotte, USA}
\affil[2]{\normalsize Department of Electrical Engineering, Alexandria University, Egypt}
\affil[3]{\normalsize Electronics and Communications Engineering Department, American University in Cairo, Egypt}
\affil[4]{\normalsize Electrical Engineering Department, Princeton University, USA}

\maketitle

\begin{abstract}
The effects of {\it quantization} and {\it coding} on the estimation quality of a Gauss-Markov, namely Ornstein-Uhlenbeck, process are considered. Samples are acquired from the process, quantized, and then encoded for transmission using either {\it infinite incremental redundancy} or {\it fixed redundancy} coding schemes. A fixed {\it processing} time is consumed at the receiver for decoding and sending feedback to the transmitter. Decoded messages are used to construct a minimum mean square error (MMSE) estimate of the process as a function of time. This is shown to be an increasing functional of the {\it age-of-information}, defined as the time elapsed since the sampling time pertaining to the latest successfully decoded message. Such (age-penalty) functional depends on the quantization bits, codeword lengths and receiver processing time. The goal, for each coding scheme, is to optimize sampling times such that the long term average MMSE is minimized. This is then characterized in the setting of {\it general increasing age-penalty functionals,} not necessarily corresponding to MMSE, which may be of independent interest in other contexts.
\end{abstract}

\section{Introduction}

Recent works have drawn connections between remote estimation of a time-varying process and the {\it age-of-information} (AoI) metric, which assesses the timeliness and freshness of the estimated data. While most works focus on transmitting {\it analog} samples for the purpose of estimation, this work focuses on using {\it quantized} and {\it coded} samples in that regard. We present optimal sampling methods that minimize the long term average minimum mean square error (MMSE) of a Gauss-Markov, namely Ornstein-Uhlenbeck (OU), process under specific coding schemes, taking into consideration receiver {\it processing} times consumed in decoding and sending feedback. 

AoI, or merely age, is defined as the time elapsed since the latest useful data has reached its destination. An increasing number of works have used AoI as a latency performance metric in various contexts, including queuing analysis, scheduling in networks and optimization under different constraints, see, e.g., the general body of works in \cite{yates_age_1, ephremides_age_random, yates_age_eh,  ephremides_age_management, ephremides_age_non_linear, modiano-age-bc, sun-age-mdp, jing-age-online, himanshu-age-source-coding, baknina-updt-info, zhou-age-iot, yates-age-mltpl-src, zhang-arafa-aoi-pricing-wiopt, batu-aoi-multihop, bacinoglu-aoi-eh-finite-gnrl-pnlty, sun-cyr-aoi-non-linear, leng-aoi-eh-cog-radio, bedewy-aoi-multihop, talak-aoi-delay, arafa-aoi-compute, inoue-aoi-general-formula-fcfs, arafa-age-online-finite, yang-arafa-aoi-fl, zou-waiting-aoi, soysal-aoi-gg11, tang-aoi-power-multi-state}.


Of particular relevance to this paper are the works related to coding for AoI improvement, e.g., \cite{parag-age-coding, najm-age-mg11-harq, yates-age-erase-code, baknina-age-coding, ceran-age-harq, sac-age-mg1-harq, simeone-age-finite-code, feng-age-rateless-codes, chen-aoi-coding-bc, arafa-aoi-coding, wang-aoi-coding-fbit, feng-coding-aoi-bc, najm-age-erasure-coding, javani-aoi-erasure}, and those related to the (inherent) role of AoI in remote estimation, e.g., \cite{chakravorty-distortion-gauss-markov, gao-estimation-ltd-measurements, yun-monitoring-comm-cost, ayan-aoi-voi-cntrl, mitra-estimation-graphs-aoi, chakravorty-estimation-pckt-drop-markov, sun-weiner, ornee-aoi-estimation-ou, huang-estimation-harq-control, maatouk-aoii, ramirez-aoi-compression, bastopcu-aoi-distortion, bastopcu-partial-updates}. Two main takeaway points from these works are: (1) optimal codes should strike a balance between using long codewords to combat channel errors and using short ones to minimize age; and (2) optimal sampling strategies should balance the need for frequent transmissions to minimize age as it relates to the correlation of the updates being received. 


\begin{figure}[t]
\center
\includegraphics[scale=.325]{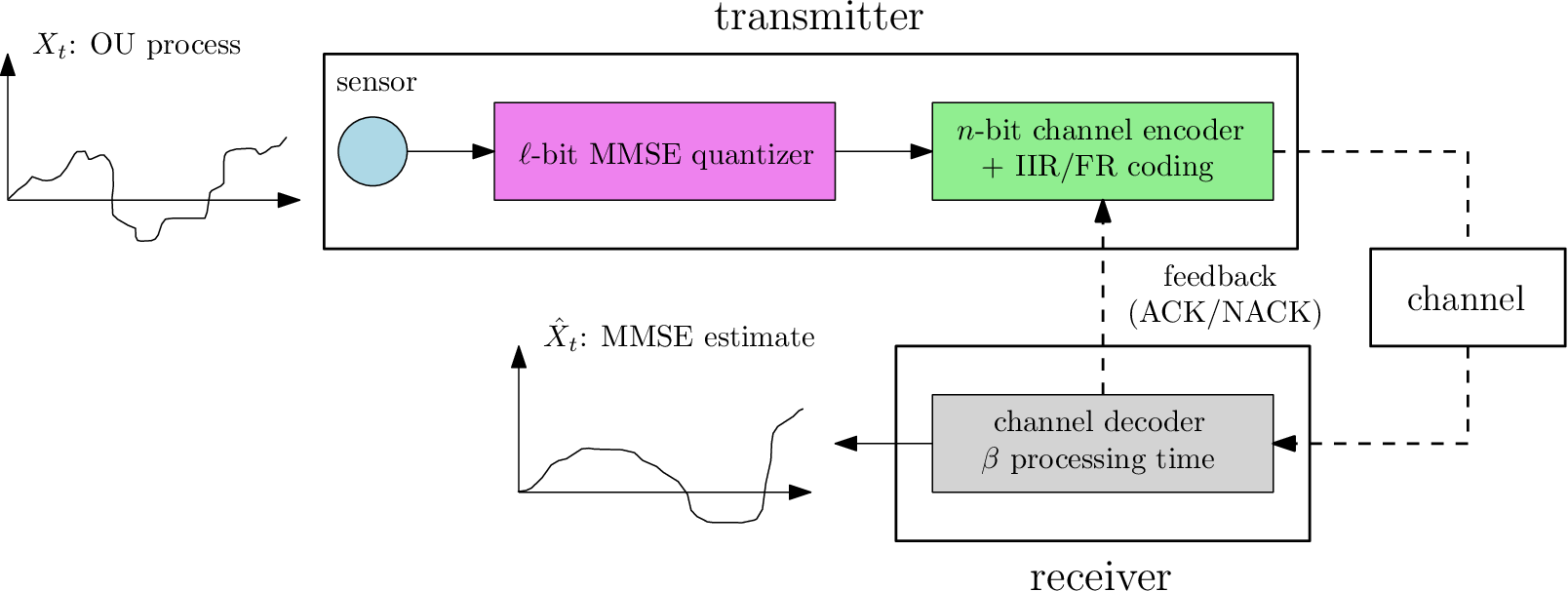}
\caption{System model considered for sampling, quantizing and encoding an OU process at the transmitter, and reconstructing it at the receiver.}
\label{fig_sys_mod}
\end{figure}

The most closely-related works to ours are \cite{sun-weiner, ornee-aoi-estimation-ou}, which derive optimal sampling methods to minimize the long term average MMSE for Weiner \cite{sun-weiner} and OU \cite{ornee-aoi-estimation-ou} processes. In both works, the communication channel introduces random delays, before perfect (distortion-free) samples are received. It is shown that if sampling times are independent of the instantaneous values of the process (signal-independent sampling) the MMSE reduces to AoI in case of Weiner \cite{sun-weiner}, and to an increasing functional of AoI (age-penalty) in case of OU \cite{ornee-aoi-estimation-ou}. It is then shown that the optimal sampling policy has a threshold structure, in which a new sample is acquired only if the expected AoI in case of Weiner (or age-penalty in case of OU) surpasses a certain value. In addition, signal-dependent optimal sampling policies are also derived \cite{sun-weiner, ornee-aoi-estimation-ou}.

In this work, we consider the transmission of quantized and coded samples of an OU process through a noisy channel.\footnote{We note that we consider an OU process in our study since, unlike the conventional Weiner process, it has a bounded variance, leading to bounded quantization error as well. The OU process, in addition, is used to model various physical phenomena, and has relevant applications in control and finance (see, e.g., the discussion in \cite{ornee-aoi-estimation-ou}).} Different from \cite{ornee-aoi-estimation-ou}, not every sample has guaranteed reception, and received samples suffer from quantization errors. The receiver uses the received samples to construct an MMSE estimate for the OU process. Quantization and coding introduce a tradeoff: {\it few quantization levels and codeword bits would transmit samples faster, yet with high distortion and probability of error.} An optimal choice, therefore, needs to be made, which depends mainly on how fast the OU process varies as well as the channel errors. Different from related works, effects of having (fixed) {\it non-zero receiver processing times}, mainly due to decoding and sending feedback, are also considered in this work.


We focus on signal-independent sampling, together with an MMSE quantizer, combined with either {\it infinite incremental redundancy} (IIR) or {\it fixed redundancy} (FR) coding schemes; see Fig.~\ref{fig_sys_mod}. The MMSE of the OU process is first shown to be an increasing functional of AoI, as in \cite{ornee-aoi-estimation-ou}, parameterized directly by the number of quantization bits $\ell$, and indirectly by the number of codeword bits $n$ and the receiver processing time $\beta$. We formulate two problems, one for IIR and another for FR, to choose sampling times so that the long term average MMSE is minimized. Focusing on stationary deterministic policies, we present optimal solutions for both problems in the case of {\it general increasing age-penalties,} not necessarily corresponding to MMSE, which may be useful in other contexts in which IIR and FR coding schemes are employed. The solution for IIR has a {\it threshold} structure, as in \cite{sun-cyr-aoi-non-linear, ornee-aoi-estimation-ou}, while that for FR is a {\it just-in-time} sampling policy that does not require receiver feedback. We finally discuss how to select $\ell$ and $n$, and show that the relatively simpler FR scheme can outperform IIR for relatively large values of $\beta$.

\section{System Model and MMSE Estimation}

We consider a sensor that acquires time-stamped samples from an OU process. Given a value of $X_s$ at time $s$, the OU process evolves as follows \cite{ou-brownian-motion, doob-brownian-motion}:
\begin{align} \label{eq_ou_evol}
X_t=X_se^{-\theta(t-s)}+\frac{\sigma}{\sqrt{2\theta}}e^{-\theta(t-s)}W_{e^{2\theta(t-s)}-1},\quad t\geq s,
\end{align}
where $W_t$ denotes a Weiner process, while $\theta>0$ and $\sigma>0$ are fixed parameters. The sensor acquires the $i$th sample at time $S_i$ and feeds it to an MMSE quantizer that produces an $\ell$-bit message ready for encoding. We will use the term {\it message} to refer to a quantized sample of the OU process. Let $\tilde{X}_{S_i}$ represent the quantized version of the sample $X_{S_i}$, and let $Q_{S_i}$ denote the corresponding quantization error. Thus,
\begin{align} \label{eq_qntz_smpl}
X_{S_i}=\tilde{X}_{S_i}+Q_{S_i}.
\end{align}
Each message is encoded and sent over a noisy channel to the receiver. The receiver updates an MMSE estimate of the OU process if decoding is successful. ACKs and NACKs are fed back following each decoding attempt, which is assumed to consume fixed $\beta\geq0$ time units. Channel errors are independent and identically distributed (i.i.d.) across time/messages.

%

Two channel coding schemes are investigated. The first is IIR, in which a message transmission starts with an $n$-bit codeword, $n\geq\ell$, and then incremental redundancy (IR) bits are added one-by-one if a NACK is received until the message is eventually decoded and an ACK is fed back. The second scheme is FR, in which a message is encoded into fixed $n$-bit codewords, yet following a NACK the message in transmission is discarded and a {\it new} sample is acquired and used instead. Following ACKs, the transmitter may idly wait before acquiring a new sample and sending a new message.

Let $D_i$ denote the reception time of the $i$th {\it successfully decoded} message. For the IIR scheme, each message is eventually decoded, and therefore
\begin{align}
D_i=S_i+Y_i
\end{align}
for some random variable $Y_i$ that represents the channel delay incurred due to the IR bits added. Let $T_b$ denote the time units consumed per bit transmission. Hence,
\begin{align}
Y_i=nT_b+\beta+r_i(T_b+\beta),
\end{align}
where $r_i\in\{0,1,2,\dots\}$ denotes the number of IR bits used until the $i$th message is decoded. Note that in the IIR scheme $\beta$ is added for the original $n$-bit codeword transmission, and then for each IR transmission until successful decoding. Let 
\begin{align}
\bar{n}\triangleq nT_b+\beta
\end{align}
for conciseness. Channel delays $Y_i$'s are i.i.d. $\sim Y$, where
\begin{align}
\mathbb{P}\left(Y=\bar{n}\right)=&p_0, \\
\mathbb{P}\left(Y=\bar{n}+k(T_b+\beta)\right)=&\prod_{j=0}^{k-1}(1-p_j)p_k,\quad k\geq1,
\end{align}
with $p_j$ denoting the probability that an ACK is received when $r_i=j$. This depends on the channel code being used, and the model of the channel errors, yet it holds that $p_j\leq p_{j+1}$.

For the FR scheme, there can possibly be a number of transmission {\it attempts} before a message is eventually decoded. Let $M_i$ denote the number of these attempts in between the $(i-1)$th and $i$th successfully decoded messages, and let $S_{i,j}$ denote the sampling time pertaining to the $j$th attempt of which, $1\leq j\leq M_i$. Therefore, only the $M_i$th message is successfully decoded, and the rest are all discarded. Since each message is encoded using fixed $n$-bit codewords, we have
\begin{align}
D_i=S_{i,M_i}+\bar{n}, \quad\forall i.
\end{align}
Observe that in the FR scheme each successfully-decoded message incurs only {\it one} $\beta$, since each decoding attempt occurs on a message pertaining to a {\it fresh} sample. According to the notation developed for the IIR channel delays above, $M_i$'s are i.i.d. geometric random variables with parameter $p_0$.

Upon successfully decoding a message at time $D_i$, the receiver constructs an MMSE estimate for the OU process. We restrict our attention to MMSE estimators that only use the latest-received information. For the IIR scheme this is
\begin{align}
\hat{X}_t=\mathbb{E}\left[X_t\Big|S_i,\tilde{X}_{S_i}\right],\quad D_i\leq t<D_{i+1}.
\end{align}
Using (\ref{eq_ou_evol}) and (\ref{eq_qntz_smpl}), we have
\begin{align}
\hat{X}_t
=&\mathbb{E}\bigg[\tilde{X}_{S_i}e^{-\theta
\left(t-S_i\right)}+Q_{S_i}e^{-\theta\left(t-S_i\right)} \nonumber \\
&\quad+\frac{\sigma}{\sqrt{2\theta}}e^{-\theta\left(t-S_i\right)}W_{e^{2\theta\left(t-S_i\right)}-1}\bigg|S_i,\tilde{X}_{S_i}\bigg] \\
=&\tilde{X}_{S_i}e^{-\theta\left(t-S_i\right)},\quad D_i\leq t<D_{i+1},
\end{align}
where the last equality follows by independence of the Weiner noise in $[D_i,t]$ from $(S_i,\tilde{X}_{S_i})$, and that for the MMSE quantizer, the quantization error is zero-mean \cite{cover}. The MMSE is now given as follows for $D_i\leq t<D_{i+1}$: (see also \cite{ornee-aoi-estimation-ou})
\begin{align}
\texttt{mse}&\left(t,S_i\right)=\mathbb{E}\left[\left(X_t-\hat{X}_t\right)^2\right] \\
=&\mathbb{E}\left[Q_{S_i}^2\right]e^{-2\theta\left(t-S_i\right)}+\frac{\sigma^2}{2\theta}\left(1-e^{-2\theta\left(t-S_i\right)}\right). \label{eq_mse_qnt_dly}
\end{align}
{\it We see from the above that even if $D_i-S_i=0$, i.e., if the $i$th sample is transmitted and received instantaneously, the MMSE estimate at $t=D_i$ would still suffer from quantization errors.}

In the sequel, we consider $X_0=0$ without loss of generality, and hence, using (\ref{eq_ou_evol}), the variance of $X_t$ is given by $\mathbb{E}\left[X_t^2\right]=\frac{\sigma^2}{2\theta}\left(1-e^{-2\theta t}\right),~t>0$.
Following a rate-distortion approach (note that $X_t$ is Gaussian), the following relates the number of bits $\ell$ and the mean square quantization error \cite{cover}:
\begin{align} \label{eq_quant_error_t}
\mathbb{E}\left[Q_t^2\right]=\frac{\sigma^2}{2\theta}\left(1-e^{-2\theta t}\right)2^{-2\ell}, \quad t>0.
\end{align}
Using the above in (\ref{eq_mse_qnt_dly}) and rearranging, we get that
\begin{align}
\!\!\!\texttt{mse}\!\left(t,S_i\right)\!=&\frac{\sigma^2}{2\theta}\!\left(\!1\!-\!\left(1\!-\!2^{-2\ell}\!\left(1\!-\!e^{-2\theta S_i}\right)\right)\!e^{-2\theta\left(t-S_i\right)}\!\right),
\end{align}
We note that as $\ell\rightarrow\infty$, the above expression becomes the same as that derived for the signal-independent sampling scheme analyzed in \cite{ornee-aoi-estimation-ou}. However, since we consider practical coding aspects in this work, as $\ell\rightarrow\infty$, it holds that $n\rightarrow\infty$ as well and no sample will be received.

We focus on dealing with the system in {\it steady state,} in which both $t$ and $S_i$ are relatively large. In this case, the mean square quantization error in (\ref{eq_quant_error_t}) becomes independent of time, and only dependent upon the steady state variance of the OU process $\sigma^2/2\theta$. Hence, in steady state, the MMSE becomes
\begin{align}
\texttt{mse}\left(t,S_i\right)=&\frac{\sigma^2}{2\theta}\left(1-\left(1-2^{-2\ell}\right)e^{-2\theta\left(t-S_i\right)}\right) \\
\triangleq&h_\ell\left(t-S_i\right), \quad D_i\leq t<D_{i+1}, \label{eq_mmse_iir}
\end{align}
which is an increasing functional of the AoI $t-S_i$.

For the FR scheme, the analysis follows similarly, after adding one more random variable denoting the number of transmissions, $M_i$. Specifically, it holds that
\begin{align}
\hat{X}_t=&\tilde{X}_{S_{i,M_i}}e^{-\theta\left(t-S_{i,M_i}\right)}, \\
\texttt{mse}\left(t,S_{i,M_i}\right)=&h_\ell\left(t-S_{i,M_i}\right), \quad D_i\leq t<D_{i+1}.
\end{align}

\section{Problem Formulation: General Age-Penalty}

The main goal is to choose the sampling times, for given $\ell$, $n$ and $\beta$, such that the long term average MMSE is minimized. In this section, we formulate two problems to achieve such goal: one for IIR and another for FR, and present their solutions in the two upcoming sections. Then, in Section~\ref{sec_cmpr_iir_fr}, we discuss how to choose the best $\ell$ and $n$, as well as compare the performances of IIR and FR in general. For both schemes, let us denote by an {\it epoch} the time elapsed in between two successfully received messages. Thus, the $i$th epoch starts at $D_{i-1}$ and ends at $D_i$, with $D_0\equiv0$.

\begin{remark}
Our analysis does not depend on the specific structure of the MMSE functional $h_\ell(\cdot)$; it extends to any differentiable increasing age-penalty functional $g(\cdot)$. Therefore, in what follows, we formulate our problems and present their solutions for the case of minimizing a long term average age-penalty, making the results applicable in other contexts.
\end{remark}

For the IIR scheme, the problem is formulated as
\begin{align} \label{opt_main_iir}
\min_{\{S_i\}}\quad\limsup_{l\rightarrow\infty}\frac{\sum_{i=0}^l\mathbb{E}\left[\int_{D_i}^{D_{i+1}}g\left(t-S_i\right)dt\right]}{\sum_{i=0}^l\mathbb{E}\left[D_{i+1}-D_i\right]},
\end{align}
where the numerator represents the total age-penalty (the MMSE in case of the OU process estimation) across all epochs, and the denominator represents the total time.


Let us define $W_i$ as the waiting time at the beginning of the $i$th epoch before acquiring the $i$th sample. That is, $S_i=D_{i-1}+W_i$. Therefore, one can equivalently solve for the waiting times $W_i$'s instead of sampling times $S_i$'s. We focus on a class of {\it stationary deterministic} policies in which
\begin{align}
W_i=f\left(g\left(D_{i-1}-S_{i-1}\right)\right),\quad\forall i.
\end{align}
That is, {\it the waiting time in the $i$th epoch is a deterministic function of its starting age-penalty value.} Such focus is motivated by the fact that channel errors are i.i.d. and by its optimality in similar frameworks, e.g., \cite{sun-age-mdp, jing-age-online, arafa-age-online-finite}. Defining $w\triangleq f\circ g$ and noting that $D_{i-1}-S_{i-1}=Y_{i-1}$ we have
\begin{align}
W_i=w\left(Y_{i-1}\right),
\end{align}
which induces a stationary distribution of $D_i$'s and the age-penalty across all epochs. Let us now (re)define notations used in a typical epoch. It starts at time $\overline{D}$ with AoI $\overline{Y}$, and with the latest sample acquired at time $\overline{S}$, such that $\overline{D}=\overline{S}+\overline{Y}$. Then, a waiting time of $w\left(\overline{Y}\right)$ follows, after which a new sample is acquired, quantized, and transmitted, taking $Y$ time units to reach the receiver at time $D=\overline{D}+w\left(\overline{Y}\right)+Y$, which is the epoch's end time. Therefore, problem (\ref{opt_main_iir}) now reduces to a minimization over a single epoch as follows:
\begin{align} \label{opt_iir_epoch}
\min_{w(\cdot)\geq0}\quad\frac{\mathbb{E}\left[\int_{\overline{D}}^{\overline{D}+w\left(\overline{Y}\right)+Y}g\left(t-\overline{S}\right)dt\right]}{\mathbb{E}\left[w\left(\overline{Y}\right)+Y\right]}.
\end{align}
Given the realization of $\overline{Y}$ at time $\overline{D}$, the transmitter decides on the waiting time $w\left(\overline{Y}\right)$ that minimizes the long term average age-penalty demonstrated in the fractional program above. We solve problem (\ref{opt_iir_epoch}) in Section~\ref{sec_iir}.

For the FR scheme, the formulated problem can be derived similarly, {\it with the addition of possible waiting times in between retransmissions.}\footnote{This is only amenable for FR since waiting leads to acquiring a fresher sample, and possibly reduced age-penalties. For IIR, waiting after a NACK is clearly suboptimal since it elongates the channel delay for the {\it same} sample.}
Specifically, let $W_{i,j}$ represent the waiting time before the $j$th transmission attempt in the $i$th epoch. A stationary deterministic policy here is such that
\begin{align}
W_{i,1}=&f\left(g\left(D_{i-1}-S_{i-1,M_{i-1}}\right)\right)=w\left(\bar{n}\right)\equiv w_1, \\
W_{i,2}=&w\left(w_1+\bar{n}\right)\equiv w_2, \\
\vdots& \nonumber \\
W_{i,j}=&w\left(w_1+\dots+w_{j-1}+\bar{n}\right)\equiv w_j,
\end{align}
and so on. Therefore, under the FR scheme, a stationary deterministic policy reduces to a countable sequence $\{w_j\}$.

Proceeding with the same notations for a given epoch as in the IIR scheme, let us define $M$ as the number of transmission attempts in the epoch, $\bar{M}$ as those in the previous epoch, and $\overline{S}_{\bar{M}}$ as the sampling time of the successful (last) transmission attempt in the previous epoch. The problem now becomes
\begin{align} \label{opt_fr_epoch}
\min_{\{w_j\geq0\}} \quad \frac{\mathbb{E}\left[\int_{\overline{D}}^{\overline{D}+\sum_{j=1}^Mw_j+M\bar{n}}g\left(t-\overline{S}_{\bar{M}}\right)dt\right]}{\mathbb{E}\left[\sum_{j=1}^Mw_j+M\bar{n}\right]}.
\end{align}
We solve problem (\ref{opt_fr_epoch}) in Section~\ref{sec_fr}.

\section{The IIR Scheme: Solution of Problem (\ref{opt_iir_epoch})} \label{sec_iir}

We follow Dinkelbach's approach to transform (\ref{opt_iir_epoch}) into the following auxiliary problem for fixed $\lambda\geq0$ \cite{dinkelbach-fractional-prog}:
\begin{align} \label{opt_iir_aux}
p^{IIR}(\lambda)\triangleq\min_{w(\cdot)\geq0}\quad&\mathbb{E}\left[\int_{\overline{D}}^{\overline{D}+w\left(\overline{Y}\right)+Y}g\left(t-\overline{S}\right)dt\right] \nonumber \\
&\hspace{.75in}-\lambda\mathbb{E}\left[w\left(\overline{Y}\right)+Y\right].
\end{align}
The optimal solution of (\ref{opt_iir_epoch}) is then given by $\lambda^*_{IIR}$ that solves $p^{IIR}(\lambda^*_{IIR})=0$, which can be found via bisection, since $p^{IIR}(\lambda)$ is decreasing \cite{dinkelbach-fractional-prog}. We now have the following result:

\begin{theorem} \label{thm_iir_main_result}
The optimal solution of problem (\ref{opt_iir_aux}) is given by
\begin{align}
w^*(\bar{y})=\left[G_{\bar{y}}^{-1}(\lambda)\right]^+,
\end{align}
where $\left[\cdot\right]^+\triangleq\max(\cdot,0)$, $\bar{y}$ is the realization of the starting AoI $\bar{Y}$, and $G_{\bar{y}}(x)\triangleq\mathbb{E}\left[g\left(\bar{y}+x+Y\right)\right]$.
\end{theorem}

Theorem~\ref{thm_iir_main_result} can be proved using \cite[Theorem~1]{sun-cyr-aoi-non-linear}. We note, however, that its proof can be approached differently from that of \cite[Theorem~1]{sun-cyr-aoi-non-linear}, in a way similar to the proof of Theorem~\ref{thm_fr_main_result} below. We omit such details due to space limits.

The optimal waiting policy for IIR has a {\it threshold} structure: a new sample is acquired only when the expected age-penalty by the end of the epoch is at least $\lambda$. Note that the optimal $\lambda^*_{IIR}$ corresponds to the optimal long term average age-penalty.


\section{The FR Scheme: Solution of Problem (\ref{opt_fr_epoch})} \label{sec_fr}

We follow a similar approach here as in the IIR scheme and consider the following auxiliary problem:
\begin{align} \label{opt_fr_aux}
&p^{FR}(\lambda)\!\triangleq\!\min_{\{w_j\geq0\}} \mathbb{E}\left[\int_{\overline{D}}^{\overline{D}+\sum_{j=1}^Mw_j+M\bar{n}}g\left(t-\overline{S}_{\bar{M}}\right)dt\right] \nonumber \\
&\hspace{1.5in}-\lambda\mathbb{E}\left[\sum_{j=1}^Mw_j+M\bar{n}\right].
\end{align}
The optimal solution of problem (\ref{opt_fr_epoch}) is now given by $\lambda^*_{FR}$ that solves $p^{FR}\left(\lambda^*_{FR}\right)=0$, which we will provide in {\it closed-form.} The optimal waiting policy structure is provided next.

\begin{theorem} \label{thm_fr_main_result}
The optimal solution of problem (\ref{opt_fr_aux}) is given by
\begin{align}
w_1^*=&\left[G^{-1}(\lambda)\right]^+, \\
w_j^*=&0,~j\geq2,
\end{align}
where $G(x)\triangleq\mathbb{E}\left[g\left(\bar{n}+x+M\bar{n}\right)\right]$. In addition, the optimal solution of problem (\ref{opt_fr_epoch}), $\lambda^*_{FR}$, is such that $w_1^*=0$.
\end{theorem}
\begin{Proof}
We first simplify the terms of the objective function of (\ref{opt_fr_aux}). Using iterated expectations, it can be shown that
\begin{align}
\mathbb{E}\left[\sum_{j=1}^Mw_j+M\bar{n}\right]=\sum_{j=1}^\infty w_j(1-p_0)^{j-1}+\frac{\bar{n}}{p_0}.
\end{align}
Now let us define
\begin{align}
\zeta_m\left({\bm w}_1^m\right)\triangleq\int_{\overline{D}}^{\overline{D}+\sum_{j=1}^mw_j+m\bar{n}}g\left(t-\overline{S}_{\bar{M}}\right)dt
\end{align}
and, leveraging iterated expectations on the first term of (\ref{opt_fr_aux}), introduce the following Lagrangian:\footnote{Using monotonicity of $g(\cdot)$, it can be shown that problem (\ref{opt_fr_aux}) is convex.}
\begin{align}
\mathcal{L}=&\sum_{m=1}^\infty \zeta_m\left({\bm w}_1^m\right)(1-p_0)^{m-1}p_0-\lambda\sum_{j=1}^\infty w_j(1-p_0)^{j-1} \nonumber \\
&-\lambda\frac{\bar{n}}{p_0}-\sum_{j=1}^\infty w_j\eta_j,
\end{align}
where $\eta_j$'s are Lagrange multipliers. Now observe that, using Leibniz rule, it holds for $j\leq m$ that
\begin{align}
\frac{\partial \zeta_m\left({\bm w}_1^m\right)}{\partial w_j}=g\left(\bar{n}+\sum_{j=1}^mw_j+m\bar{n}\right).
\end{align}
Taking derivative of the Lagrangian with respect to $w_j$ and equating to $0$, we use the above to get
\begin{align} \label{eq_fr_pf_wj}
\sum_{m=j}^\infty g\!\left(\!\bar{n}+\sum_{j=1}^mw_j+m\bar{n}\!\right)\!(1-p_0)^{m-j}p_0\!=\lambda\!+\!\frac{\eta_j}{(1\!-\!p_0)^{j-1}}.
\end{align}
Next, let us substitute $j=k$ and $j=k+1$ above, $k\geq1$, subtract them from each other, and rearrange to get
\begin{align}
g\left(\bar{n}+\sum_{j=1}^kw_j+k\bar{n}\right)=\lambda+\frac{\eta_k-\eta_{k+1}}{(1-p_0)^{k-1}p_0}.
\end{align}
Since $g(\cdot)$ is increasing, and $\lambda$ is fixed, $\left\{\frac{\eta_k-\eta_{k+1}}{(1-p_0)^{k-1}p_0}\right\}$ is increasing. From there, one can conclude that $\eta_j>0,~j\geq2$ must hold. Hence, by complementary slackness, $w_j^*=0,~j\geq2$ \cite{boyd}. Using (\ref{eq_fr_pf_wj}) for $j=1$, the optimal $w_1^*$ now solves
\begin{align}
G\left(w_1^*\right)=\lambda+\eta_1,
\end{align}
where $G(\cdot)$ is as defined in the theorem. Observe that $G(\cdot)$ is increasing and therefore the above has a unique solution. Now, if $\lambda\leq G(0)$, then we must have $\eta_1>0$, and hence $w_1^*=0$ by complementary slackness; conversely, if $\lambda>G(0)$, then we must have $w_1^*>0$, and hence $\eta_1=0$ by complementary slackness as well \cite{boyd}. In the latter case, $w_1^*=G^{-1}(\lambda)$. Finally, observe that $\lambda\leq G(0)\iff G^{-1}(\lambda)\leq0$. This concludes the proof of the first part of the theorem.

To show the second part, all we need to prove now is that $G^{-1}\left(\lambda^*_{FR}\right)\leq0$, or equivalently that $\lambda^*_{FR}\leq G(0)$. Toward that end, observe that $p_{FR}(\lambda)$ is decreasing, and therefore if $p_{FR}\left(G(0)\right)\leq0$ then the premise follows. Now for $\lambda=G(0)$ we know from the first part of the proof that $w_1^*=0$. Thus,
\begin{align}
p_{FR}&\left(G(0)\right)=\sum_{m=1}^\infty\zeta_m\left(0\right)(1-p_0)^{m-1}p_0-G(0)\frac{\bar{n}}{p_0} \\
=&\mathbb{E}\left[\int_{\overline{D}}^{\overline{D}+M\bar{n}}g\left(t-\overline{S}_{\bar{M}}\right)dt\right]-G(0)\mathbb{E}\left[M\right]\bar{n} \\
=&\mathbb{E}\left[\int_{0}^{M\bar{n}}g\left(\bar{n}+t\right)dt\right]-\mathbb{E}\left[\int_0^{M\bar{n}}G(0)dt\right] \label{eq_fr_pf_w1_0_1} \\
=&\mathbb{E}\left[\int_{0}^{M\bar{n}}\mathbb{E}\left[g\left(\bar{n}+t\right)-g\left(\bar{n}+M\bar{n}\right)\right]dt\right],\label{eq_fr_pf_w1_0_2}
\end{align}	
where (\ref{eq_fr_pf_w1_0_1}) follows by change of variables and (\ref{eq_fr_pf_w1_0_2}) follows by definition of $G(\cdot)$. Finally, observe that by monotonicity of $g(\cdot)$, (\ref{eq_fr_pf_w1_0_2}) is non-positive. This concludes the proof.
\end{Proof}

A closed-form expression for $\lambda^*_{FR}$ can now be found via substituting $w_j=0,~\forall j$ in (\ref{opt_fr_epoch}).

Theorem~\ref{thm_fr_main_result} shows that {\it zero-wait} policies are optimal for FR, which is quite intuitive. First, waiting is not optimal in between retransmissions, even though it would lead to acquiring fresher samples, since the AoI is already relatively high following failures. Second, since the epoch always starts with the same AoI, $\bar{n}$, one can optimize the long term average age-penalty to make waiting not optimal at the beginning of the epoch as well. We note, however, that such results do {\it not} follow from \cite[Theorem~5]{sun-age-mdp}, since there can be multiple transmissions in the same epoch. We also note that while zero-wait policies have been invoked in other works involving FR coding schemes, e.g., \cite{yates-age-erase-code, najm-age-erasure-coding}, Theorem~\ref{thm_fr_main_result} provides a proof of their {\it optimality} for general increasing age-penalties.

Now that zero-waiting is optimal, we investigate whether it is necessary to wait for the receiver to decode and send feedback before sending the next message. It could be better, age-wise, to send a new message right away after the previous one is {\it delivered}, i.e., after $nT_b$ time units instead of $\bar{n}$. However, this may not be optimal if $\beta$ is relatively large, since it would lead to accumulating {\it stale} messages at the receiver's end as they wait for decoding to finish. The next lemma revolves this. The proof is similar, in essence, to that of Theorem~\ref{thm_fr_main_result}, and is omitted due to space limits.

\begin{lemma} \label{thm_fr_waiting}
In the FR scheme, a new message is sent after the previous one's delivery by $\left[\beta-nT_b\right]^+$ time units.
\end{lemma}

%
%

Lemma~\ref{thm_fr_waiting} shows that {\it just-in-time} updating is optimal. For $\beta\leq nT_b$, a new sample is acquired and transmitted just-in-time as the previous message is delivered. While for $\beta>nT_b$, a new sample is acquired and transmitted such that it is delivered just-in-time as the receiver finishes decoding the previous message. This way, delivered samples are always fresh, the receiver is never idle, and feedback is unnecessary.

\section{Discussion: Which Scheme Performs Better?} \label{sec_cmpr_iir_fr}

We now discuss the IIR and FR performances in the original context of OU estimation, i.e., when $g(\cdot)\equiv h_\ell(\cdot)$. Applying Theorem~\ref{thm_iir_main_result}'s result, the optimal waiting policy for IIR is
\begin{align}
w^*\!\left(\bar{y}\right)\!=\!\left[\frac{1}{2\theta}\log\left(\frac{\frac{\sigma^2}{2\theta}\left(1-2^{-2\ell}\right)\mathbb{E}\left[e^{-2\theta Y}\right]}{\frac{\sigma^2}{2\theta}-\lambda^*_{IIR}}\right)-\bar{y}\right]^+.
\end{align}
One can also show that $\lambda^*_{IIR}\in\left[2^{-2\ell}\frac{\sigma^2}{2\theta},\frac{\sigma^2}{2\theta}\right]$, facilitating the bisection search. Applying Theorem~\ref{thm_fr_main_result} and Lemma~\ref{thm_fr_waiting}'s results, the optimal long term average MMSE for FR is given by
\begin{align}
\frac{\sigma^2}{2\theta}\left(\!1\!-\!\frac{\left(1-2^{-2\ell}\right)e^{-2\theta\bar{n}}p_0}{2\theta K_{n,\beta}}\frac{1-e^{-2\theta K_{n,\beta}}}{1\!-\!(1-p_0)e^{-2\theta K_{n,\beta}}}\right),
\end{align}
where $K_{n,\beta}\triangleq\max\{\beta,nT_b\}$. Derivation details are omitted.

We consider a binary symmetric channel with crossover probability $\epsilon\in\left(0,\frac{1}{2}\right)$, and use MDS codes for transmission. This allows us to write $p_j=\sum_{l=0}^{\lfloor\frac{n+j-\ell}{2}\rfloor}\binom{n+j}{l}\epsilon^l(1-\epsilon)^{n+j-l}$, where $\lfloor x\rfloor$ denotes the highest integer not larger than $x$. We set $\sigma^2=1$, and $T_b=0.05$ time units. For fixed $\beta=0.15$, we vary $\ell$ and choose the best $n$ for IIR and FR via grid search, and then choose the best allover $\ell$. We repeat for $\theta=0.01$ (slow process variation) and $\theta=0.5$ (fast process variation). The results are shown in Table~\ref{tab_ell}. We see that the optimal $\ell^*$ that minimizes the MMSE is inversely proportional with $\theta$. This is intuitive, since for slowly varying processes, one can tolerate larger waiting times to get high quality estimates, and vice versa. Similar behavior is noticed for relatively good ($\epsilon=0.1$) and bad ($\epsilon=0.4$) channel conditions.

\begin{figure}[t]
\center
\includegraphics[scale=.375]{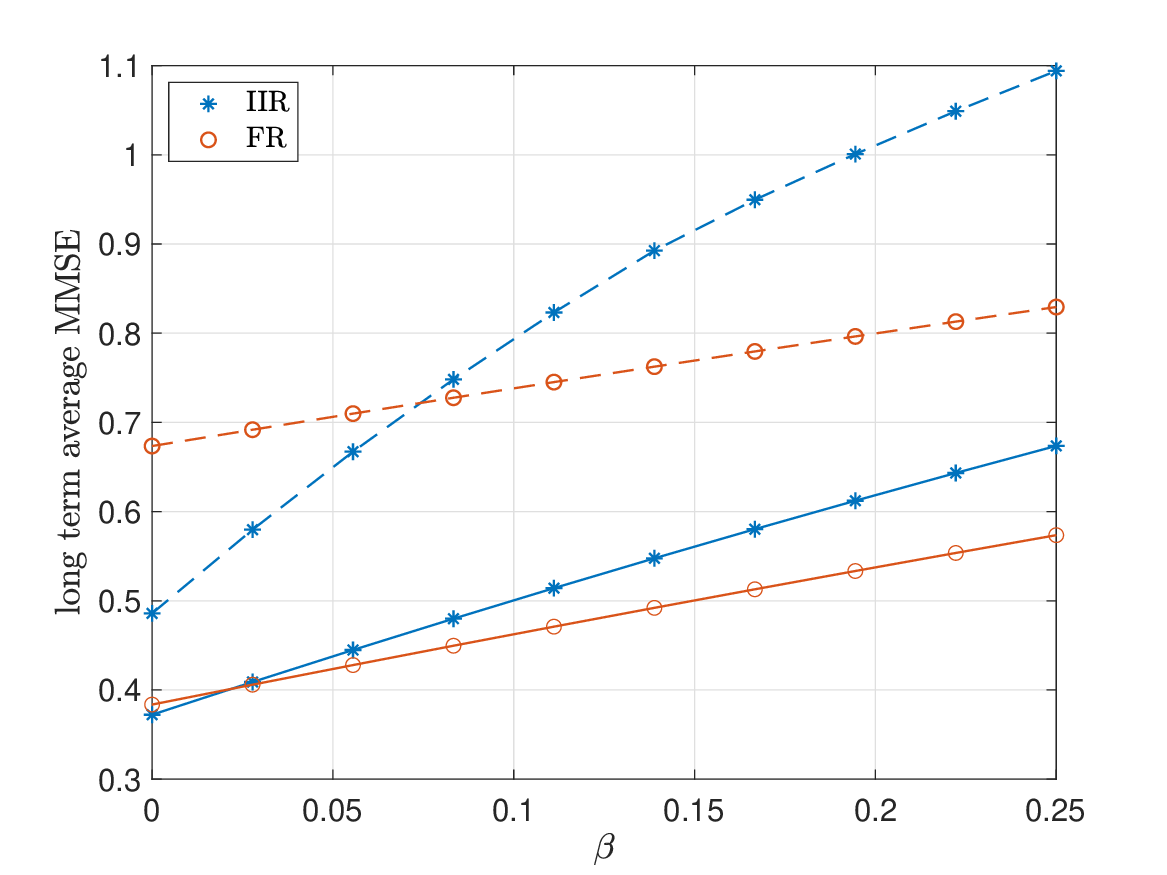}
\vspace{-.1in}
\caption{$\ell=3$; $\theta=0.25$; dashed lines: $\epsilon=0.4$, and solid lines: $\epsilon=0.1$.}
\label{fig_iir_vs_fr_beta}
\vspace{-.275in}
\end{figure}

In Fig.~\ref{fig_iir_vs_fr_beta}, we fix $\theta=0.25$, $\ell=3$ bits and vary $\beta$. We see that FR outperforms IIR for relatively large $\beta$. This is clearer in worse channel conditions ($\epsilon=0.4$), which is expected since the $\beta$ processing penalty is incurred, following NACKs, at the bit level for IIR and at the message level for FR.

\begin{table}
\center
{\footnotesize 
\begin{tabular}[t]{l*{6}{c}r}
Setting & $\epsilon=0.1$: & IIR & FR & $\epsilon=0.4$: & IIR & FR\\
\hline
$\theta=0.01$ & & $(5,7)$ & $(5,7)$ & & $(4,10)$ & $(4,6)$\\
$\theta=0.5$ & & $(2,4)$ & $(2,4)$ & & $(1,3)$ & $(2,4)$
\end{tabular}
\caption{Optimal $(\ell^*,n^*)$ for IIR and FR under different settings.}
\label{tab_ell}
\vspace{-.2in}
}
\end{table}




\end{document}